\documentclass [12pt]{article}
\usepackage {amssymb}
\usepackage {amsmath}

\begin{document}

\begin{center}
\textbf{ON SPACETIME DIFFERENTIAL ELEMENTS AND THE DISTRIBUTION OF
BIO-HAMILTONIAN COMPONENTS}
\end{center}

\begin{center}
Michel Bounias
\end{center}

\begin{center}
University of Avignon, Faculty of Sciences and INRA-DSPE,
Biomathematics Unit Domain of Sagne-Soulier, \\F-07470 Le Lac
d'lssarl\`{e}s, France
\end{center}

\textbf{Abstract.} Various Hamiltonian models have been derived
for chemical structures belonging to living organisms while the
Hamiltonian concept was not applied to life as a whole. However,
Hamiltonian components were recently defined for living organisms
on the condition to take in consideration their evolutionary
implications (Bounias, 2001: CASYS'0l). This paper identifies
differential elements of Spacetime, from which it delimits a
probabilistic fuzzy-like invariance standing for conservativity of
biological Hamiltonians. The distributions of potential and
kinetic components in a individual bio-Hamiltonian, and the
distribution of such individual Hamiltonians of living organisms
interacting in more complex systems are shown to behave as a
non-linear generalized convolution of functions.



\bigskip

\section*{Introduction}

\hspace*{\parindent} While living organisms do not behave
independently from the properties of matter (Bounias, 1990), for
long, no Hamiltonian, nor wave function nor Schr\"odinger equation
was considered for living systems (Rosen, 1989). The concept of a
Hamiltonian of a system was originally defined for physical
systems in classical and quantum mechanics, them for simple
chemical systems. In the recent past years, Hamiltonian treatment
has been tried for components of living organisms. Structures were
addressed in proteins, for solitons in Raman scattering
(Xiao-Feng, 1998) and beta sheet to alpha helix conformations
(Ito, 1999), in DNA helix-coil transition (Morozov et al., 2000),
in plant light-harvesting chromoprotein complexes (Tretiak et al.,
2000). Functions were considered in electron transfer tunneling
(Balabin et al., 1998), and energy storage for cellular motion
(Nakagawa et al., 2000). All such works have been dealing with
Hamiltonian treatment of structures involved in living organisms,
that is concerning chemical molecules rather than the living
phenomenon in its whole.

However, while Hamiltonian and wave equations are used in Physics
to try to predict the evolution of a system, up to the evolution
of universe, if similar parameters were to be identified for
living organisms, they would contribute to predict the behavior of
ecosystems in connection with the status of their embeding medium,
namely Planet Earth. The main components of the Hamiltonian of
life have recently been shown to include: (i) kinetic components
as the manifold WK=\{WK$_{\rm m}$, WK$_{\rm M}$, WK$_{\rm E}$\} of
microstructural and metabolic interactions, macroscopic activity
and anticipatory behavior leading to homeostatic and evolutionary
adaptation; (ii) potential components WP=\{WP$_{\rm m}$, WP$_{\rm
M}$, WP$_{\rm E}$\}$\cup \left( {\rm WP}_{\rm g} \right)$, the
latter including the selection of expressed characters from DNA
existing structures, and the construction of new genomic
components by evolutionary processes (Bounias, 2001).

Since living organisms are interacting in more complex systems and
ecosystems, where they are embedded it was necessary to examine by which
kind of relations their respective Hamiltonians, which may be only partly
conservative, could themselves be connected within more conservative
supersystems. This study will address first the distribution of kinetic and
potential components of an individual Hamiltonian throughout the
time-related sequence of configurations, and then the distribution of
interacting Hamiltonians inside a more complex system.

\bigskip

\section{On differential elements of spacetime}

\hspace*{\parindent} Former works have demonstrated that our
observable spacetime can be formally identified with a ordered
sequence \{S$_{{\kern 1pt}\rm i}$\} of 3-D Poincar\'{e} sections
embedded in a 4-D topological space (Bonaly and Bounias, 1995).
Mappings of one into the next section wear the form of a momentum
and stand for infinitesimal increments of time and space (Bounias,
1997). The embedding topological 4-space is provided with a
natural metrics as the set distance, i.e. the symmetric
differences between sets (Bounias and Bonaly, 1996; Bounias,
1997), which is compatible with the definition of a topology on a
space. Each section is mapped to the next one by a moment of
junction (MJ) which connects either the distances or the objects,
i.e. their complementaries or "instans" (Bounias, 1997). In short,
space is subdivided into sets intersections, standing for objects
(or "instans") denoted by m $<$A, B, ...$>$ and their
complementaries, i.e. the set distances denoted by $\Delta ({\rm
A,B,...})$.

\textbf{Definition 1.} The Moments of Junction are defined as follows for G
= (m or $\Delta $) and X=\{A,B,...\}:

 $$
{\rm MJ_{G( {i,j})}\quad = \quad G_i[ {X} ] \bot {\kern 1pt}{\kern
1pt}f_{( {i,j} )} ( {X} )} \eqno(1)
 $$

\noindent where function \textit{f} takes values 0$ \leqslant
f_{\left( {\rm i,j} \right)} $(X)$ \leqslant $l (Bounias, 1997),
depending on the indicatrix functions l(x) of each point x of a
section (S$_{{\kern 1pt}\rm i}$) mapped into the topologies of the
next (S$_{{\kern 1pt}\rm i+1}$), or generally to any further one
(S$_{{\kern 1pt}\rm j}$). For any closed and open subparts P$_{\rm
i}$(X) in (S$_{{\kern 1pt}\rm i}$), one has for any x:

\medskip

$$1_{\rm i} \left( {x} \right) = \left| {{\begin{array}{*{20}c}
{1\quad {\rm iff}{\kern 1pt} \;\,{\rm x} \in \left( {\rm P_{i}}
\right),} \hfill \\ {0\quad {\rm iff}{\kern 1pt} \;\,{\rm x}
\notin \left( {\rm P_{i}}  \right),} \hfill \\
\end{array}} } \right.$$

\noindent then:

$$f_{(\rm i,j)} \left( {x} \right) = \left|
{{\begin{array}{*{20}c} {1\quad {\rm iff}{\kern 1pt} \;\,1_{\rm
i}({\rm x}) =1_{\rm j}({\rm x}), } \hfill
\\ {0\quad {\rm iff}{\kern 1pt} \;\, 1_{\rm i}({\rm x}) \neq 1_{\rm
j}({\rm x}).} \hfill \\
\end{array}} } \right.\eqno(2)$$

\medskip

\textbf{Theorem 1.} The Moment of Junction provides a differential
element of spacetime.

Proof. Let a space increments from (S$_{{\kern 1pt}\rm i }$) to
(S$_{{\kern 1pt}\rm i+1}$) be as small as a difference in one
point. Thus, for the mapping of (S$_{{\kern 1pt}\rm i }$) into
(S$_{\rm i+1}$) the Moment of Junction MJ$_{({\kern 1pt}\rm
i,i+1)}$ differs by a distance defined by $\rm d(x^\prime_i,
{\kern 1pt} x_{i+1})$ where $\rm x^\prime_i$ is the projection of
$\rm x_i$ on (S$_{\rm i+1}$). Two such points can be adjacent
though nonequal, that is the distance $\rm d(x^\prime_i, {\kern
1pt} x_{i+1})$ can be as small as needed, while MJ$_{({\kern
1pt}\rm i,i+1)}$ remains the same mathematical object. Therefore,
MJ$_{({\kern 1pt}\rm i,i+1)}$ stands for a differential element of
space.

Then, as far as there exists at least one point $\rm x_i$ such
that $\rm d(x^\prime_i, {\kern 1pt} x_{i+1})$, then $\rm
(S)\cap(S)\neq\varnothing$ and the Moment of Junction is positive.
Hence, MJ$_{({\kern 1pt}\rm i,i+1)}$ represents in this case the
smallest interval separating two states of the considered space.
This interval exists, it is non-null, though it has no measurable
duration. This denotes a differential element of time.

Gathered together, these two statements define a differential of both space
and time, that is of spacetime, which completes the proof.

\section{Interaction mappings of bio-Hamiltonian components}

\subsection{Distribution of components of individual Hamiltonians}

\textbf{Lemma 2.1.1.} The moment of junction of the Hamiltonian of
a conservative system is distributive for its components.

Proof. Let W denote the kinetic component and V the potential one
in H=(W+V). A variation $\rm (W-dW)$ is accompanied by a
correlated (V+dV). In the spacetime sequence, $f_{\rm (.i)}{\rm
(X_i-dX_i)}\mapsto f_{\rm (.j)}{\rm (X_j-dX_i)}=f_{\rm (.j)}{\rm
(X_j) + f_{\rm (.j)}(dX_i)}$ for objects composing the set X in
which W and V can ultimately be measured. Then:
\begin{eqnarray*}
{\rm  MJ(W\cup V)} &=& {\rm MJ[(W \backslash dW) \cup (V\cup dV)]}
\\   &=& {\rm MJ(W)\cup MJ(V) \cup (dV \backslash dW)]}
\end{eqnarray*}
  $$
\rm{ with \ \ \rm (dV\backslash dW)=\oslash \ \ iff \ \
dV=dW}.\quad\quad \quad\quad\quad \quad\quad\quad\quad
\quad\quad\quad \quad\quad\quad\quad\quad \quad\quad
  $$
(Note that the denotation A$\backslash$B above signifies the
complementary of B in A.)

  Thus:
   $$
\rm MJ(W \cup V) = MJ(W) \cup MJ(V)
   $$
iff the  system  is  conservative.

 \textbf{Lemma 2.1.2.}
The Hamiltonian of a individual organism is affected a boundary of
invariance.

Proof. Let H(W,V) be the Hamiltonian of an organism A$\in$(X) and
$\varphi$ a function such that: $\rm MJ(W\cup V) =
\varphi(MJ\{A\})$. Then, $\rm H(W\cup V) = \varphi(H\{A\})$.
Assuming that the system A is measured by continuous variables,
the moment MJ of $\varphi$(W,V) can be written using the joined
probability density of W and V, i.e. f(W,V) (Ruegg, 1988):
  $$ \rm
MJ[\varphi(W,V)] = \int \int  \varphi (W,V) f(W,V) dW dV \eqno(3)
  $$
Assume the particular case where $\varphi$(W,V) = W$\cup$V. Then:
  $$
\rm MJ(W \cup V) = \int \int (W \cup V) f(W,V) dW dV \eqno(4)
  $$
The repartition function of ${\rm H = W }\cup {\rm V}$ is F(h),
for ${\rm H}=\{ {\rm h_1,\ ...\ , h_n }\}$ is:

  $$ {\rm F(h)} = \int\int\limits_{{\rm W} \cup {\rm V} \subseteq
{\rm h} \quad\quad\quad} {\it f}\rm (W) {\kern 1pt} {\it f} \rm
(V){\kern 1pt}{\kern 1pt} dW{\kern 1pt} dV \eqno(5)
  $$

\noindent where h appears as a boundary delimiting the range of
invariance of H.

\textbf{Remarks.}

(i) The distribution function $f_{\rm (i,j)}$(A) is valued in
[0,1] and such is valued the distribution of components giving the
measure of W and V. Therefore, the invariance boundary introduces
the notion of a fuzzy invariance for the Hamiltonian of a
biological organism whose components are provided a apparent
stability by flows of matter and energy from exchanges with the
surrounding milieu.

(ii) Function $f_{\rm (i,j)}$(A) defines the balance of system (A)
between W and V forms:

At extrema of global values, $f_{\rm (i,j)}(\rm A)=1$ denotes a
absolutely motionless state (W=0) while $f_{\rm (i,j)}(\rm A)=0$
depicts a state of absolute motion (V=0).

\subsection{Distributions of Hamiltonian functions for two
interacting organisms}

\textbf{Definitions 2.2.1.} Denote by X=\{A,B,Q\} the set of
species, habitat and resources, respectively. The global ecosystem
is a space of magmas [4] E=\{(X), ($\Phi$)\}, where ($\phi$) is a
functional. Call (O) and ($\perp$) two kinds of mappings
connecting Hamiltonians H(x$_{\rm i}$) and H(x$_{\rm j}$) for any
two members of (X) and (T$^\perp$) the family of mappings from
($\perp$) to some (O). Call ($\varphi$) the specific kind of
relationship which maps two components H(x$_{\rm i}$) and
H(x$_{\rm j}$) contained in H\{(x$_{\rm i}$), (x$_{\rm j}$)\}. Let
H[(x$_{\rm i}$) $\cup$ (x$_{\rm j}$)]$\mapsto \varphi$[H(x$_{\rm
i}$), H(x$_{\rm j}$)] be a function (approximated as H(x$_{\rm
i}$) $\cup$ H(x$_{\rm j}$) in section 3.1). Note that dH(x$_{\rm
i}) \neq 0$, dH(x$_{\rm i})\neq 0$ during interaction, with
dH(x$_{\rm i}$, x$_{\rm j}\approx 0$ for $\varphi(x_{\rm i},
x_{\rm j}) \subseteq \{\rm x_{\rm i},\rm x_{\rm j}\} \subseteq
(\rm X) \subseteq (\rm E)\subseteq (\rm  etc.)$.

Repartition functions still are denoted by F and distribution
functions by $f$.

\textbf{Theorem 2.2.2.} Hamiltonians of individual components of a
invariant pair in a system with higher order of complexity are
mapped by non-linear convolution-like functions.

Proof. For continued variables, let H(z) = $\varphi$H(x$_{\rm
i}$), H(x$_{\rm j}$)). Then:
  $$
{\rm F(H(z)} = \int\limits_{\varphi \rm (H(x_{i}),H(x_{j}))
\subseteq H\{x_{i},x_{j}\} )} f(\rm H(x_i), H(x_{j})){\kern
1pt}{\kern 1pt}dH(x_{i}) \cdot dH(x_{j}) \eqno(6)
  $$

\noindent where H\{x$_{\rm i}$, x$_{\rm j}$\} stands for the
former fuzzy invariant boundary h of relation (4).

For discrete variables one would have the following distribution of
probabilities:

  $$ \rm P\big( \varphi(H(x_{i}), H(x_{j})) \big)
   = \mathop{\cup}\limits^{\rm k\in (X)}
_ {\rm k\in (\oslash)} P\big\{ \big( \varphi \rm
(H(x_{i})=k \big)
  $$
  $$
\cap \big( \varphi \rm (H(x_{j}) \big) = \complement_{\rm x}
\varphi \rm (H(x_{j})) \} \quad\quad\quad\quad\quad\quad\quad
\eqno(7)
  $$

\noindent where $\complement_{\rm A}({\rm B})$ denotes the
complementary of B in A, also denoted by A$\backslash$B.

Reducing relations (6) and (7) to the particular case where one
would have: $\varphi$(H(x$_{\rm i}$), H(z))=(H(x$_{\rm i}$) +
(x$_{\rm j}$)) would give for a discrete variable:

  $$
\rm P(H(x_{i}) + H(x_{j})) = \sum\limits^{\rm Hz}_{\rm k=0} \{ \rm
P(Hx_{i}=k) \cap \rm (Hx_{j}=Hz-k)\} \eqno(8)
  $$

\noindent and for a continuous variable the repartition function:

  $$
\rm F(Hz_{i})=\int\limits^{+\infty}_{-\infty} {\it f} \rm (Hx_{i})
\cdot F(Hz-Hx_{i}){\kern 1pt} dHx_{i} \eqno(9)
  $$

\noindent that is also the distribution, with commutativity
between Hx$_{\rm i}$ and Hx$_{\rm j}$:

  $$
f(\rm Hz_{i})=\int\limits^{+\infty}_{-\infty} {\it f}\rm (Hx_{i})
\cdot {\it f} \rm (Hz-Hx_{i}){\kern 1pt}dHx_{i} \eqno(10)
  $$

\noindent which denotes the convolution $f(\rm Hx_{i}) \ast {\it
f} (Hx_{j})$.

This allows an extension of the general case of the functional
($\Phi$). In effect: \ let i and j be indexed on Card(X), k be
indexed on a spatial distribution within any of Poincar\`{e}
sections (S$_\alpha$) of the ordered sequence \{S\}$_n$, and L be
indexed on the sequence ($n\in$ L). Then, the mappings of ($\Phi$)
are involved in the following two expressions:

  $$
\big( \rm (Hx_{\rm i}) \perp^{\rm L} (Hx_{j}) \big)_{L+t} =
T_{L}^\perp \big( (Hx_{i}) {\kern 1pt} O^{\rm L}
(Hx_{j})\big)_{\rm L}, \eqno(11a)
  $$
  $$
\big( \rm (Hx_{\rm i}) \perp^{\rm k} (Hx_{j}) \big)_{k+p} =
T_{k}^\perp \big( (Hx_{i}) {\kern 1pt} O^{\rm k}
(Hx_{j})\big)_{\rm k},   \eqno(11b)
  $$

that is, by gathering (11a) and (11b) into one single form:

  $$
\big( \rm (Hx_{\rm i}) \perp^{\rm L \ast k} (Hx_{j})
\big)_{(L+t)\ast (k+p)} = T_{L\ast k}^\perp \big( (Hx_{i}) {\kern
1pt} O^{\rm L \ast k} (Hx_{j})\big)_{\rm L \ast k}, \eqno(12)
  $$

\noindent which denotes a nonlinear generalized convolution in the
sense of Bolivar-Toledo et al. (1985).

\begin{flushright}
(QED)
\end{flushright}

\subsection{Boundaries of the system}

\hspace*{\parindent} Now, some preliminary consideration should be
added about the area of validity of the above functionals.

\textbf{Definitions 2.3.1.} We will call "canonic functions" the conditions
for the functionality of ecosystems which apply to all members as
equivalence relations or in a commutative way (which includes the Abelian
groups for all binary relations operating with relevant kinds of mappings).
Examples are the founding conditions (Bonaly and Bounias, 2000) of
continuity, complementarity and mutualism.

We will call "specific functions" those which connect interspecific
relations as order relations. An example is the relation "feeding on" in
predator-to-prey relations.

\textbf{Proposition} 2.3.2. The domain of the convolution of
Hamiltonians [equation (13)] belongs to the set of canonical
functions, and its range belongs to the complete system of
canonical plus specific functions.

How specific functions are involved will be matter of further developments.

All these results provide a perspective for further exploration of
relationships connecting Hamiltonian components of the Hamiltonian
of a global system.

\section{Discussion and Conclusion}

\subsection{Outside components in potential and kinetic energies}

\hspace*{\parindent} The bio-Hamiltonian has been shown to be
under influence of external factors, though it represents an
internal sum of energy. A potential energy W$_{\rm P}$ or  is the
product of a scalar $\mu$ (characteristic of components of mass of
an object) by a distance of functions $\rm
d[\zeta(x_{i}),\zeta(x_{j})]$ of its positions, where $\zeta$ maps
a causality factor applying on $\mu$. It is noteworthy that
E$_{\rm Pot}$ of a system involves the work that forces (i.e.
causality components) acting on a system are able to perform,
taking into account the parameters of position, shape,
configuration, of this system. Thus, components outside the system
are involved.

The kinetic energy w$_{\rm k}$ or E$_{\rm kin}$ is a function of
some expression of the mass M of a system (M=$\cup {\kern 1pt}\rm
m_{i}$) and of the square of the velocities $\rm (v_{i})^2$ of its
components, in a Newtonian, a relativistic and related forms.
Importantly, the theorem of the kinetic energy states that the
variation of kinetic energy of a system during a time lapse is the
sum of all works of all forces (i.e. causality parameters) acting
on the system during this interval, thus including internal,
external and connection or interaction forces. Since $\rm v_{i} =
d x_{i} /d t$, the position of objects is again involved.

\subsection{Wave function for macroscopic objects}

\hspace*{\parindent} In classical quantum mechanics, the wave
function $\psi$ is determined by the frequency $\nu$ and by the de
Broglie wavelength ($\lambda_{\rm de{\kern 1pt}Broglie}$) of a
particle (Krasnoholovets, 2001b). So far, no physical
interpretation was possible for $\psi$ as the root of a
probability of localization. However, recently the wave function
of a macroscopic object has alternatively been shown to be
conceivable in terms of specific deformations of space, by
Krasnoholovets (2001a,b). The period and amplitude of a system
composed of a peculiar form of deformation of space (standing for
a particle whose mass is proportional to the deformations)
periodically communicated partly to the surrounding space (giving
a "inerton cloud") and then back to the particle. During this
cycle, the velocity of a moving particle oscillates between an
initial value and zero, and its mass components oscillate between
the particle and its inertons cloud (Krasnoholovets, 1997).

This approach provides a physical meaning to the de Broglie and
Compton wavelengths as well as to the frequency of the system, and
the corresponding formalism has been shown to reach a classical
form. Let \{$\pi$\} be a set of vector parameters describing all
of the mass components of the corpuscular system and ${\hat
c}_{\kern 1pt \pi}$ a limit in the velocity of transmission of
space deformations; then, $\ddot \pi - {\hat c}^{\kern 1pt
2}_{\kern 1pt \pi} \nabla \pi = 0$ (Krasnoholovets, 2002). Wave
function components of one particle can thus be extended to those
of an entire organism and to all massive objects. Furthermore, the
theory consistently allows gravity and relativity to be deduced
from submicroscopic properties (Krasnoholovets, 1997, 2000,
2001a). Therefore, a deterministic macroscopic wave function
$\psi$(X,t) becomes conceptually accessible and it can be
associated with the Hamiltonian of living organisms. In a
preliminary work (Bounias, 2001) it has been pointed out that the
trace of the macroscopic wave function of a ecosystem in the
sequence \{S$_{\kern 1pt i}$\} of Poincar\`{e} sections stands for
the historical of the ecosystem, a non-linear causality factor
identified by Landis (1996).

\subsection{Specific conservativity status \\ of the
bio-Hamiltonian}

\hspace*{\parindent} Studying the Hamiltonian of a living organism
rather than just biochemical components raises a property of
fuzzy-like conservativity which contrasts with the status of
physical objects. However, no physical structure is strictly
conservative: the ceaseless motion does not exist, and all
corpuscles have limited duration of life. In a molecule, atoms
have different Hamiltonians, and the Hamiltonian of the molecule
itself is subjected to the nature of interactions with its
environment.

In a more complex system like a ecosystem, all components of
individual Hamiltonians are interacting in a dynamical steady
state. It has been demonstrated (Bounias and Bonaly, 2000) that
the state of such an ecosystem is determined by the properties of
the orbit of each component (which includes species, habitat and
resources) by the manifold of functions. All combinations of these
parameters are timely non-linear and the evolution of the system
is logically determined by a non-linear convolution: this supports
the result obtained here from a more fundamental approach
involving the moments of junction as differential elements of
spacetime.

The fuzzy-invariance component appearing in biological systems
represents a term with topological meaning. In effect, the
convolution of bio-Hamiltonians correlates all their components in
a compact space since it is finite and discrete. The
Heine-Borel-Lebesgue theorem states that a finite subcover can
exist from any finite subcover: the latter is necessarily finite
and it involves all possible correlations, of which some actually
are reflected in a finite section of spacetime. This lets a choice
about which components are selected in a redundant system as Life,
and therefore the presence of a fuzzy operator is justified. On
the other hand, while the invariance of moments originates in
empirical observations, and remains to be formally proved from a
completely independent theory, conservativity has been shown to be
fulfilled through a continuum of the geometry of physical objects
in a 4-manifold, where only their traces in 3-D sections have a
physical meaning.

\bigskip

\textbf{Acknowledgements}

\medskip

 The author wishes to thank Dr. Volodymyr
Krasnoholovets for stimulating discussions and for providing supportive
materials to this research.

\bigskip

\noindent {\bf References}

\medskip

{\small

\noindent Balabin, L.A., Onuchic, J.N., 1998. A new framework for
electron-transfer calculations beyond the pathways-like models.
{\it J. Phys. Chem. B} {\bf 102}, 7497-7505.

\vspace{2mm}

\noindent Bolivar-Toledo, 0., Candela, S., Munoz-Blanco, J.A.,
1985. Non linear data transforms in perceptual systems. In:
"Lecture Notes in Computer Science", G. Goos and J. Hartmanis
eds., No. 410, 1-9.

\vspace{2mm}

\noindent Bonaly, A., Bounias, M., 1995. The trace of time in
Poincar\`{e} sections of a topological space. {\it Phys. Essays},
8(2), 236-244.

\vspace{2mm}

\noindent Bounias, M., 1990. "La cr\'{e}ation de la vie: de la
mati\`{e}re \`{a} 1'esprit". Ed. du Rocher, Paris, 444 pp.

\vspace{2mm}

\noindent Bounias, M., Bonaly, A., 1996. On metrics and scaling:
physical coordinates in topological spaces. {\it Ind. J. Theor.
Phys.}, 44(4), 303-321.

\vspace{2mm}

\noindent Bounias, M., 1997. Definition and some properties of
set-difference, instans and their momentum, in the search for
probationary spaces. {\it Ultra Scientist of Phys. Sciences},
9(2), 139-145.

\vspace{2mm}

\noindent Bounias, M., Bonaly, A., 2000. "The future of life on
Earth: ecosystems as topological spaces". In: "The Future of the
Universe and The Future of Our Civilization". V. Burdyuzha \& G.
Khozin eds., Proc. 1st.

\vspace{2mm}

\noindent Bounias, M., 2002. The Hamiltonian of life: an
anticipatory operator of evolution. In: CASYS'Ol Int. Math. Conf.,
ed. by D. Dubois, Li\`{e}ge (Belgium), Aug. 13-18, 2001. {\it Int.
J. Comput. Anticipatory Syst.}, in press.

\vspace{2mm}

\noindent Ito, H., 1999. Probability tensor theory based upon a
Bariey-Nielsen-Schellman-type model Hamiltonian: circular
dichroism calculations of polypeptides. {\it J. Chem. Phys.},
111(19), 9093-9110.

\vspace{2mm}

\noindent Krasnoholovets, V., 1997. Motion of a relativistic
particle and the vacuum. {\it Phys. Essays}, 10(3), 407- 416 (also
quant-ph/9903077).

\vspace{2mm}

\noindent Krasnoholovets, V., 2000a. On the nature of spin,
inertia and gravity of a moving canonical particle. {\it Ind. J.
Theor. Phys.} 48(2), 97-132 (also quant-ph/0103110).

\vspace{2mm}

\noindent Krasnoholovets, V., 2000b. Space structure and quantum
mechanics. {\it Spacetime \& Substance} 1(4) 172-175 (also
quant-ph/0106106).

\vspace{2mm}

\noindent Krasnoholovets, V., 2001a. On the way to submicroscopic
description of nature. {\it Ind. J. Theor. Phys.}, 49(2), 81-95
(also quant-ph/9906091).

\vspace{2mm}

\noindent Krasnoholovets, V., 2001b. Submicroscopic deterministic
quantum mechanics, in: CASYS'2001 Int. Math. Conf., Daniel M.
Dubois ed., Li\`{e}ge (Belgium), 13-18 Aug. 2001. {Int. J. Comput.
Anticip. Systems} (2002), in press (also quant-ph/0109012)

\vspace{2mm}

\noindent Krasnoholovets, V., 2002. Gravitation as deduced from
submicroscopic quantum mechanics, submitted.

\vspace{2mm}

\noindent Landis, W.G., 1996. The integration of environmental
toxicology. {\it SETAC News}, March 1996, 15-16.

\vspace{2mm}

\noindent Morozov, V.F., Mamasakhlisov, E.S., Hayryan, S., Hu,
C.K., 2000. Microscopical approach to the helix-coil transition in
DNA. {\it Physica A} (Amsterdam), 281(1-4), 51-59.

\vspace{2mm}

\noindent Nakagawa, N., Kaneko, K., 2000. Energy storage in a
Hamiltonian system in partial contact with a heat {\it J. Phys.
Soc. Jpn,} 69(5), 1255-1258.

\vspace{2mm}

\noindent Rosen, R., 1989. On socio-biological homologies, 28 pp.
Some epistemological issues in physics and biology (dedicated to
David Bohm), 28pp. Undated papers kindly communicated to the
author.

\vspace{2mm}

\noindent Ruegg, 1985. Probabilities and Statistics. Presses
universitaires Romandes, Lausanne (Suisse).

\vspace{2mm}

\noindent Tretiak, S., Middleton, C., Chernyak, V., Mukamel, S.,
2000. Frenkel exciton Hamiltonian for LH2 photosynthetic antenna.
Polym. Prepr. (Am. Chem. Soc., Div. Polym. Chem.), 41(1), 879-880.

\vspace{2mm}

\noindent Xiao-Feng, P., 1998. The properties of Raman scattering
resulting from solitons excited in the organic protein molecules.
{\it Acta Phys. Slovacia}, 48(2), 99-114.

\end{document}